\documentclass[12pt,preprint]{aastex}
\usepackage{textcomp}
\usepackage{amsmath}
\usepackage{graphicx}
\usepackage[T1]{fontenc}
\shorttitle{Global solutions of NDAFs}
\shortauthors{Xue et al.} \slugcomment{}

\def\beq{\begin{eqnarray}}
\def\eeq{\end{eqnarray}}

\begin{document}

\title{Relativistic global solutions of neutrino-dominated accretion flows}

\author{Li Xue\altaffilmark{1,2}, Tong Liu\altaffilmark{1,3}, Wei-Min Gu\altaffilmark{1}, and Ju-Fu Lu\altaffilmark{1}}

\altaffiltext{1}{Department of Astronomy and Institute of Theoretical Physics and Astrophysics, Xiamen University, Xiamen, Fujian 361005, China; tongliu@xmu.edu.cn}
\altaffiltext{2}{Nicolaus Copernicus Astronomical Center, Bartycka 18, 00-716 Warszawa, Poland}
\altaffiltext{3}{State Key Laboratory of Theoretical Physics, Institute of Theoretical Physics, Chinese Academy of Sciences, Beijing, 100190, China}

\begin{abstract}
Neutrino-dominated accretion flows (NDAFs) around rotating stellar-mass black holes are plausible candidates for the central engines of gamma-ray bursts (GRBs). We investigate one-dimensional global solutions of NDAFs, taking account of general relativity in Kerr metric, neutrino physics and nucleosynthesis more precisely than previous works. We calculate sixteen solutions with different characterized accretion rates and black hole spins to exhibit the radial distributions of various physical properties in NDAFs. \iffalse We find that the gas pressure and the neutrino cooling always become dominated in the inner region for large accretion rate, $\gtrsim 0.1~M_\odot$ $\rm s^{-1}$. \fi We confirm that the electron degeneracy has important effects in NDAFs and we find that the electron fraction is about $0.46$ in the outer region for all the sixteen solutions. From the perspective of the mass fraction, free nucleons, $\rm ^4He$, and $\rm ^{56}Fe$ dominate in the inner, middle, and outer region, respectively. The influence of neutrino trapping on the annihilation is of importance for the superhigh accretion ($\dot M=10M_{\odot}~\rm s^{-1}$) and most of the sixteen solutions have an adequate annihilation luminosity for GRBs.
\end{abstract}

\keywords {accretion, accretion disks - black hole physics - gamma-ray burst: general - nuclear reactions, nucleosynthesis, abundances}

\section{Introduction}

The observations of gamma-ray bursts (GRBs) are well explained by the relativistic fireball shock model to some extents. However, the central engine powering the fireball are always hidden inside resulting from extremely thick optical depth of the fireball. A popular model of central engine named neutrino dominated accretion flow (NDAF) involves a hyperaccreting stellar-mass black hole with the accretion rates in the range of $0.01 \sim 10 M_\odot~{\rm s}^{-1}$. This model has been widely applied to explain the variable light-curves, extended emission, X-ray flares, associated supernovae and gravitational radiation and so on in GRBs in the past decade \citep[e.g.,][]{Popham1999,Narayan2001,Di Matteo2002,Kohri2002,Kohri2005,Lee2005,Gu2006,Chen2007,Janiuk2007,Kawanaka2007,Liu2007,Liu2008,Liu2010a,Liu2010b, Liu2012a,Liu2012b,Liu2013,Lei2009,Romero2010,Sun2012,Kawanaka2012a,Li2013}. In this developing route of NDAF theory, much more detailed and precise microphysics has been widely introduced to improve the theory \citep[see,
e.g.,][]{Kato2008}.

The relativistic global solutions of NDAF were firstly worked out by \cite{Popham1999}. They found that the inner region of NDAF is in the extremely hot and dense state and the free electrons are in the degenerated state in which the photons are totally trapped and only neutrinos can escape to carry away the viscously dissipated gravitational energy. Those neutrinos would collide with each other and happen the annihilation of neutrino pairs in a funnel space above the inner disk of NDAF to produce a relativistic fireball of a GRB event. In their solutions, they assumed that NDAFs are always optical thin anywhere for neutrinos, even for the case with extremely high accretion rate, and they over simplified the treatment of neutrino production and electron degeneracy. These result in the overstated annihilation luminosity, especially for the high accretion rate, and lose much information of microphysics. Therefore, much subsequent research works are dedicated to improve the microphysics of NDAF \citep[e.g.,][]{Di Matteo2002,Kohri2002,Kohri2005,Lee2005,Janiuk2007}. Some elaborate physical considerations, such as the definition of neutrino optical depth, precise treatment of electron degeneracy and electron fraction, are introduced to improve the NDAF theory step by step. \citet{Gu2006} showed that general relativistic effects should be considered and the contribution from the neutrino-optically thick region should also be included. Under such consideration, they found that NDAF can still work as the central engine of GRB from the viewpoint of energy. \citet{Liu2007} studied the radial structure and neutrino annihilation luminosity of NDAF. They introduced a bridging formula to treat the radial distribution of the electron fraction between neutrino optical thin and thick limits, but they ignored the existence of heavy metal elements and assumed that the heaviest nucleus is $^4 \rm He$, which implies that the numerical value of the electron fraction at the radial outer boundary is $0.5$. \citet{Chen2007} presented calculations of the structure of NDAFs around Kerr black holes and proved that both the electron degeneracy and the electron fraction dramatically affect the structure. They also considered that $^4 \rm He$ abounded in the out region of the disk. The ignition radius and other characteristic radii are defined in their work. \citet{Kawanaka2007} investigated NDAFs around Schwarzschild black hole with pseudo-Newtonian potential \citep{Paczynski1980}. They assumed that the inflowing nucleon gas is composed primarily of nuclei of neutron-rich iron group, and the electron fraction is $0.42$ at the outer boundary. They studied the radial structure and stability of the disk for the different mass accretion rate, using a realistic equation of state \citep{Lattimer1991}, in order to properly treat the dissociation of nuclei. \citet{Kawanaka2012a} studied the effects of the convection in NDAFs. They proposed that this process can be use to explain the origin of the highly variable light-curves in the prompt emissions of GRBs. \citet{Liu2013} investigated the vertical structure and elements distribution of NDAFs in spherical coordinates with the reasonable nuclear statistical equilibrium \citep[NSE,][]{Seitenzahl2008}. According their calculations, heavy nuclei tend to be produced in a thin region near the disk surface, whose mass fractions are primarily determined by the accretion rate and the vertical distribution of temperature and density. In this thin region, they found that $^{56}\rm Ni$ is dominant for the flow with low accretion rate (e.g., $0.05 $$M_{\odot}$~$\rm s^{-1}$) but $^{56}\rm Fe$ is dominant for the high counterpart (e.g., $1 M_{\odot}$~$\rm s^{-1}$). The dominant $^{56}\rm Ni$ in the special region may provide a clue to understand the bumps in the optical light curve of core-collapse supernovae.

In this paper, we return to investigate the relativistic global solutions of NDAF in Kerr metric, but we fully upgrade the microphysical treatment with the detailed neutrino physics and precise NSE based the improvement of NDAF theory in past decade. In section \ref{sec_phy_mod}, we establish our physical model for NDAF through introducing the fundamental hydrodynamic and thermodynamic equations, the detailed neutrino processes and the proton-rich NSE \citep{Seitenzahl2008,Liu2013}. In section 3, we describe in detail the numerical methods for the calculations. In section \ref{sec_num_res}, we show some typical solutions and discuss the results revealed from them. We investigate the solutions with different characterized accretion rates and black hole spins. For each case, we calculate neutrino luminosity and neutrino annihilation luminosity, and show their dependence on these parameters. Conclusions and discussion are made in section \ref{sec_con_dis}.

\section{Physical Model}\label{sec_phy_mod}

\subsection{Relativistic hydrodynamics}
In this paper, we consider to solve for the disk structure in the Kerr metric, because the inner regions of disks may contribute most of the luminosity and it is effected deeply by the spin of the black hole. Our hydrodynamical model of disks is based on the advection-dominated accretion flow (ADAF) model of \cite{Abramowicz1996}, the NDAF model of \cite{Popham1999}, and the slim disk model of \cite{Sadowski2009}, which are all research works on the one-dimensional global solutions of accretion disks in Kerr metric. For the convenience, we describe our hydrodynamical model in the units of $G=c=M=1$ ($M$ is the mass of black hole), but we turn to use cgs units when we describe the neutrino physics and thermodynamics, and present our results later.

The continuity equation is \beq \label{continuity} \dot{M}=-4\pi\rho H \Delta^{1/2}\frac{V_r}{\sqrt{1-{V_r}^2}},\eeq where $\dot M$ is the rest-mass accretion rate, $\rho$ is the rest-mass density, $H$ is the half thickness of disk, $V_r$ is the radial velocity measured in the corotating frame, $\Delta\equiv r^2-2 r+a^2$ is a function of the Boyer-Lindquist radial coordinate $r$, and $a$ is the total specific angular momentum of the black hole.

The gas energy equation is \beq \label{energy}-\frac{\dot{M}}{2\pi r^2}\left(\frac{u}{\rho}\frac{d\ln u}{d\ln r}-\frac{p}{\rho}\frac{d\ln\rho}{d\ln r}\right)=-\frac{2\alpha p H A \gamma^2}{r^3}\frac{d\Omega}{dr}-Q^-,\eeq where $u$ is the specific internal energy, $p$ is the pressure, $\alpha$ is the viscosity parameter, $A\equiv r^4+r^2 a^2+2 r a^2$, $\gamma$ is the Lorentz factor, $\Omega\equiv u^{\phi}/u^{t}$ is the angular velocity with respect to the stationary observer, and $Q^-$ is the total cooling rate as described in $\S$\ref{sec_thermodynamics}.

The radial momentum equation is \beq \label{radial-momentum} \frac{V_r}{1-{V_r}^2}\frac{d{V_r}}{dr}=\frac{\mathcal{A}}{r}-(1-{V_r}^2)\frac{1}{\lambda\rho}\frac{dp}{dr}, \eeq where \beq \mathcal{A}\equiv-\frac{A}{r^3\Delta\Omega^+_K\Omega^-_K}\frac{(\Omega-\Omega^+_K)(\Omega-\Omega^-_K)}{1-\tilde{\Omega}^2\tilde{R}^2}.\eeq The $\mathcal{A}$ term combines the effects of gravity and rotation, where $\lambda\equiv(\rho+p+u)/\rho$ is the relativistic enthalpy, $\tilde{\Omega}\equiv\Omega-2ar/A$ is the angular velocity with respect to the local inertial observer, $\Omega^{\pm}_K\equiv\pm(r^{3/2}\pm a)^{-1}$ are the angular frequencies of the corotating and counterrotating Keplerian orbits, and $\tilde{R}\equiv A/(r^2\Delta^{1/2})$ is the radius of gyration.

The equation of angular momentum conservation is \beq \label{angular-momentum} \dot{M}(\mathcal{L}-\mathcal{L}_{\rm{in}})=\frac{4\pi p H A^{1/2}\Delta^{1/2}\gamma}{r}, \eeq where $\mathcal{L}\equiv u_{\phi}$ is the specific angular momentum of the accreting gas, $\mathcal{L}_{\rm{in}}$ is the specific angular momentum at the inner edge of the disk.

The equation of vertical mechanical equilibrium \citep{Abramowicz1997} is \beq\label{vertical-eq} \frac{p}{\lambda\rho H^2}=\frac{\mathcal{L}^2-a^2(\epsilon^2-1)}{r^4}, \eeq where $\epsilon\equiv u_t$ is the energy at infinity, which is conserved along geodesics. In the practical calculation, the detailed evaluating formulae of $\epsilon$ and $\gamma$ are necessary, \beq \epsilon=-\gamma\frac{r\Delta^{1/2}}{A^{1/2}}-\frac{2ar}{A}\mathcal{L},\eeq and \beq \gamma=\sqrt{\frac{1}{1-{V_r}^2}+\frac{\mathcal{L}^2r^2}{A}}.\eeq

\subsection{Neutrino physics}

The main difference between NDAF and typical accretion disk is the cooling mechanism. Neutrino radiation becomes dominated in NDAF, so the microphysics, especially the neutrino physics must be included in the calculations.

\subsubsection{Neutrino optical depth}

The total optical depth for neutrinos is \beq \tau_{\nu_i}=\tau_{s, \nu_i} + \tau_{a,\nu_i}, \eeq where $\tau_{s,\nu_i}$ and $\tau_{a,\nu_i}$ are the neutrino optical depth from scattering and absorption, the subscript $i$ runs for the three species of neutrinos $\nu_{\rm e}$ , $\nu_{\rm \mu}$ , and $\nu_{\rm \tau}$.

The optical depth for neutrinos through scattering off electrons and nucleons $\tau_{s, \nu_i}$ is given by \beq \tau_{s,\nu_i} \approx H \displaystyle(\sigma_{{\rm e}, \nu_i} n_{\rm e} + {\sum_{j}} \sigma_{j,\nu_i} n_j), \eeq where $H$ is the half thickness of the disk, $\sigma_{{\rm e}, \nu_i}$, $\sigma_{j, \nu_i}$, $n_{\rm e}$ and $n_j$ ($j=1$, 2,...) are the cross sections of electron and nucleons ($n_1$ and $n_2$ are the number density of free protons and free neutrons), and the number density of electrons and nucleons ($j \geq 3$), respectively \citep[e.g.,][]{Kohri2005,Chen2007,Kawanaka2007,Liu2007,Liu2012a}. The major cross sections are from scattering off electrons, free protons, free neutrons and other elements particles, which are given by \citep{Burrow2004,Chen2007} \beq \sigma_{{\rm e},\nu_i}\approx \frac{3 k_B T \sigma_0 e_{\nu_i}}{8 m_{\rm e} c^2} (1+ \frac{\eta_{\rm e}}{4})[(C_{V,\nu_i}+C_{A,\nu_i})^2+ \frac{1}{3} (C_{V,\nu_i}-C_{A,\nu_i})^2],\eeq \beq \sigma_{{\rm n_1},\nu_i}\approx\frac{\sigma_0 e_{\nu_i}^2}{4}[(C_{V,\nu_i}-1)^2+3 g_A^2 (C_{A,\nu_i}-1)^2] ,\eeq \beq \sigma_{{\rm n_2},\nu_i}\approx\frac{\sigma_0 e_{\nu_i}^2}{4} \frac{1+3 g_A^2}{4}, \eeq \beq \sigma_{n_j,\nu_i}\approx \frac{\sigma_0}{16} e^2_{\nu_i} (Z_j + N_j) [1-\frac{2Z_j}{Z_j+N_j}(1-2 \sin^2 \theta_W)]^2,\eeq  where $k_B$ and $\eta_{\rm e}$ are the the Boltzmann constant and electron degeneracy, $\sigma_0= 4G^2_F(m_{\rm e}c^2)^2/\pi(\hbar c)^4 \\ \approx 1.71 \times 10^{- 44} {\rm cm^2}$, $G_F \approx 1.436 \times 10^{-49} {\rm erg~cm^3}$, $e_{\nu_i}$ is the mean energy of neutrinos in units of $(m_{\rm e} c^2)$, $g_A \approx 1.26$, $\sin^2 \theta_W \approx0.23$, $Z_j$ and $N_j$ are defined as the number of the protons and neutrons of a nucleus $X_j$, $C_{V,\nu_e}=1/2 + 2 \sin^2 \theta_W$ , $C_{V,\nu_\mu} =C_{V,\nu_\tau}= - 1/2 + 2\sin^2 \theta_W$ , $ C_{A,\nu_e}= C_{A,\overline{\nu}_\mu}=C_{A,\overline{\nu}_\tau}=1/2$, $C_{A,\overline{\nu}_e} = C_{A,\nu_\mu}=C_{A,\nu_\tau}=- 1/2$.

The number density of electrons and positron can be given by the Fermi-Dirac integration \citep[see, e.g.,][]{Kohri2005,Kawanaka2007,Liu2007}, \beq\ n_{\rm e^{\mp}}= \frac{1}{\hbar^3 \pi^2} \int_0^\infty d p~p^2 \frac{1}{e^{({\sqrt{p^2 c^2+{m_{\rm e}}^2 c^4} \mp {\mu_{\rm e}})/k_{\rm B} T}}+1}, \eeq where $\mu_{\rm e}=\eta_{\rm e} k_B T$ is the chemical potential of electrons.

The absorption depth for neutrinos $\tau_{a, \nu_i}$ is defined by \beq\ \tau_{a, \nu_i}= \frac{q_{\nu_i} H}{4 (7/8) \sigma T^4},\eeq where $q_{\nu_i}$ is the total neutrino cooling rate (per unit volume) and is the sum of four terms, \beq q_{\nu_i}=q_{\rm Urca}+q_{{\rm e^{-}+e^{+}} \rightarrow \nu_i+ \overline {\nu}_i}+q_{{\rm n + n \rightarrow  n + n +} \nu_i+\overline{\nu}_i}+q_{\tilde{\gamma} \rightarrow \nu_i+\overline \nu_i}.\eeq

Urca processes have been included in the proton-rich NSE \citep{Seitenzahl2008,Liu2013}. The neutrino cooling rate due to the Urca processes $q_{\rm Urca}$ relates only to $\nu_{\rm e}$ and for simplicity, we considered that there are four major terms by electrons, positron, free protons, free neutrons and nucleons \citep{Liu2007,Kawanaka2007}, in other words, the energy emission rate for electron capture by heavy nuclei is important for the outer region of the disk \citep{Kawanaka2007}. The main energy emission rate is \beq q_{\rm Urca}=q_{\rm p+e^{-}\rightarrow n+\nu_{\rm e}}+q_{\rm n+e^{+}\rightarrow p+\overline{\rm \nu}_{\rm e}}+q_{\rm n \rightarrow p+e^{-}+\overline{\nu}_{\rm e}}+q_{ X_j+{\rm e^{-}}\rightarrow X'_j+\nu_{\rm e}}, \eeq with \beq q_{\rm p+e^{-}\rightarrow n+\nu_{\rm e}} = \frac{G_F^2 \cos^2 \theta_c}{2 \pi^2 \hbar^3 c^2}(1+3 g_A^2) n_{\rm 1} \int_Q^\infty d E_{\rm e}~E_{\rm e}~\sqrt{{E_{\rm e}}^2-{m_{\rm e}}^2 c^4}(E_{\rm e}-Q)^3 f_{{\rm e}^{-}} ,\eeq \beq q_{\rm n+e^{+}\rightarrow p+\overline{\nu}_{\rm e}}=\frac{G_F^2 \cos^2 \theta_c}{2 \pi^2 \hbar^3 c^2}(1+3 g_A^2) n_{\rm 2} \int_{m_{\rm e} c^2}^\infty d E_{\rm e}~E_{\rm e}~\sqrt{{E_{\rm e}}^2-{m_{\rm e}}^2 c^4}(E_{\rm e}+Q)^3 f_{{\rm e}^{+}}, \eeq \beq q_{\rm n \rightarrow p+e^{-}+\overline{\nu}_{\rm e}}=\frac{G_F^2 \cos^2 \theta_c}{2 \pi^2 \hbar^3 c^2}(1+3 g_A^2) n_{\rm 2} \int_{m_{\rm e} c^2}^Q d E_{\rm e}~E_{\rm e}~\sqrt{{E_{\rm e}}^2-{m_{\rm e}}^2 c^4}(Q-E_{\rm e})^3 (1-f_{{\rm e}^{-}}), \eeq \beq q_{ X_j+{\rm e^{-}}\rightarrow X'_j+\nu_{\rm e}} = \frac{G_F^2 \cos^2 \theta_c}{2 \pi^2 \hbar^3 c^2}g_A^2 \frac{2}{7} N_p(Z_j) N_h(N_j)n_j \int_{Q'}^\infty d E_{\rm e}~E_{\rm e}~\sqrt{{E_{\rm e}}^2-{m_{\rm e}}^2 c^4}(E_{\rm e}-Q')^3 f_{{\rm e}^{-}},\eeq where $\cos^2 \theta_c \approx 0.947$, $Q = (m_{\rm n} - m_{\rm p})c^2$, $Q'\approx \mu'_{\rm n}-\mu'_{\rm p}+\Delta$, $\mu'_{\rm n}$ and $\mu'_{\rm p}$ are the chemical potential of protons and neutrons in their own nuclei, $\Delta\approx3 \rm MeV$, and $f_{{\rm e}^{\mp}}=\{\exp[(E_{\rm e} \mp \mu_{\rm e})/k_{\rm B} T]+1\}^{-1}$ is the Fermi-Dirac function. \beq N_p(Z_j) =\begin{cases} 0, ~~Z_j<20,\\Z_j-20, ~~20<Z_j<28, \\ 8, ~~Z_j>28, \end{cases}\eeq \beq N_h(N_j) =\begin{cases} 6, ~~N_j<34,\\40-N_j, ~~34<N_j<40, \\ 0, ~~N_j>40. \end{cases}\eeq The third term (also named $\beta$ decay) is small comparing with the first two terms, and was usually not included in the literature.

The electron-positron pair annihilation rate into neutrinos $q_{{\rm e^{-}+e^{+}} \rightarrow \nu_i+ \overline \nu_i}$ is \citep[e.g.,][]{Itoh1989,Yakovlev2001,Janiuk2007} \beq \begin{split} q_{{\rm e^{-}+e^{+}} \rightarrow \nu_i+ \overline {\nu}_i}= &\frac{Q_c}{36 \pi} \{(C^2_{V,\nu_i}+C^2_{A,\nu_i})^2[8(\Phi_1 U_2 + \Phi_2 U_1)-2(\Phi_{-1} U_2 + \Phi_2 U_{-1}) \\ & + 7(\Phi_0 U_1 + \Phi_1 U_0)]\}+ \{[5(\Phi_0 U_{-1} + \Phi_{-1} U_0)]\\ &+9(C^2_{V,\nu_i}-C^2_{A,\nu_i})^2 [\Phi_0(U_1+U_{-1})+(\Phi_{-1}+\Phi_1)U_0]\}, \end{split} \eeq where $Q_c = ({m_{\rm e} c}/{\hbar})^9 {G^2_F}/{\hbar} \approx 1.023 \times 10^{23} \rm erg~cm^{-3}~s^{-1}$, and the dimensionless functions $U_k$ and $\Phi_k$ ($k= -1, 0, 1, 2$) in the above equation can be expressed in terms of the Fermi-Dirac functions \citep{Kawanaka2007}. As we know, when electrons are degenerate $q_{{\rm e^{-}+e^{+}} \rightarrow \nu_i+ \overline \nu_i}$ becomes negligible.

The nucleon-nucleon bremsstrahlung rate $q_{{\rm n + n \rightarrow  n + n +} \nu_i+\overline{\nu}_i}$ is the same for the three species of neutrinos \citep[e.g.,][]{Itoh1996,Di Matteo2002,Liu2007}, \beq q_{{\rm n + n \rightarrow  n + n +} \nu_i+\overline{\nu}_i} \approx 1.5 \times 10^{27} \rho_{10}^2 T_{11}^{5.5} ~{\rm erg~cm^{-3}~s^{-1}}.\eeq

We considered that the plasmon decay rate $q_{\tilde{\gamma} \rightarrow \nu_i+\overline \nu_i}$ needs to be considered, where plasmons $\tilde{\gamma}$ are photons interacting with electrons \citep[e.g.,][]{Kawanaka2007}, \beq q_{\tilde{\gamma} \rightarrow \nu_{\rm e}+\overline \nu_{\rm e}} = \frac{\pi^4}{6 \alpha^*} C_{V,\nu_{\rm e}} \frac{\sigma_0 c}{(m_{\rm e}c^2)^2} \frac{(k_B T)^9}{(2 \pi \hbar c)^6} \gamma^6 (\gamma^2+2\gamma+2){\rm exp}(-\gamma),\eeq \beq q_{\tilde{\gamma} \rightarrow \nu_\mu+\overline \nu_\mu} = q_{\tilde{\gamma} \rightarrow \nu_\tau+\overline \nu_\tau} = \frac{4 \pi^4}{6 \alpha^*} C_{V,\nu_\mu} \frac{\sigma_0 c}{(m_{\rm e}c^2)^2} \frac{(k_B T)^9}{(2 \pi \hbar c)^6} \gamma^6 (\gamma^2+2\gamma+2){\rm exp}(-\gamma),\eeq where $\alpha^* \approx 1/137$ is the fine-structure constant, and $\gamma \approx 5.565 \times 10^{-2} {[(\pi^2 + 3 \eta_e^2)/3]}^{1/2}$. Here $q_{{\rm n + n \rightarrow  n + n +} \nu_i+\overline{\nu}_i}$ and $q_{\tilde{\gamma} \rightarrow \nu_i+\overline \nu_i}$ may become important only for very high electron degeneracy state.

\subsubsection{Electron Fraction}

The electron fraction can be written as \citep[e.g.,][]{Liu2013} \beq Y_{\rm e}=\frac{\sum\limits_{j} n_j Z_j} {\sum\limits_{j} n_j (Z_j+N_j)}. \eeq

\cite{Liu2007} calculated the electron fraction according to the simple NSE equation, the condition of electrical neutrality and a bridging formula of electron fraction that is valid in both the optically thin ($\mu_{\rm n}=\mu_{\rm p}+2\mu_{\rm e}$, where $\mu_{\rm n}$ and $\mu_{\rm p}$ are the chemical potential of free neutrons and protons) and thick ($\mu_{\rm n}=\mu_{\rm p}+\mu_{\rm e}$) regimes. Here we use the strict NSE equations (see Section 2.3) to replace the simple one which assumed that the heaviest nuclei is $^4 \rm He$. Meanwhile, the condition of electrical neutrality still holds, which can be given by \citep{Liu2007,Liu2013} \beq \sum\limits_{j} n_j Z_j=\frac {\rho Y_{\rm e}}{m_u} = n_{\rm e^{-}}-n_{\rm e^{+}},\eeq where $m_u$ is the mean mass of nucleus, and we considered that the mass fraction approximately equals the number density.

Furthermore, in order to allow for a transition from the optically thin to optically thick regimes, the bridging formula of free protons and neutrons can be established by the relations of the reaction rates in the above $\beta$
processes, which can be written as \citep{Yuan2005,Liu2007}, \beq \label{bridging-formula} \lg{\frac{n_{\rm 2}}{n_{\rm 1}}}= f(\tau_\nu) \frac{2 \mu_{\rm e}-Q}{k_{\rm B} T}+[1-f(\tau_\nu)] \frac{\mu_{\rm e}-Q}{k_{\rm B} T}, \eeq where $f(\tau_\nu)=\rm exp(-\tau_{\nu_{\rm e}})$ is a weight factor, and $\mu_{\rm e}$ is the chemical potentials of electrons. In addition, the bridging formula also can be used even taking into account nucleosynthesis because there is neutrino optical thin in the outer region of the disk.

\subsection{Nucleosynthesis}

NSE established by all nuclear reactions are in the chemical equilibrium. \cite{Seitenzahl2008} studied proton-rich material in a state of NSE, which applies to almost all the range of the electron fraction. The complicated and detailed balance has been included under the condition of the equilibrium of chemical potential. The number density of nucleus $j$ can be considered as \beq n_j = g_j (\frac{m_j k_B T}{\hbar^2})^{3/2}{\rm exp}[\frac{Z_j (\mu^{\rm kin}_{\rm p}+{\mu^{\rm C}_{\rm p}})+N_j {\mu^{\rm kin}_{\rm n}}-{\mu^{\rm C}_{\rm j}}+Q_j}{k_B T}],\eeq where $\mu^{\rm kin}_{\rm p}$ and $\mu^{\rm kin}_{\rm n}$ are the kinetic chemical potentials of protons and neutrons, $\mu^{\rm C}_{\rm p}$ and $\mu^{\rm C}_{\rm j}$ are the Coulomb chemical potentials of protons and nucleons, and $g_j$ is the nuclear partition functions. They showed that $\rm ^{56}Ni$ is favored in NSE under proton-rich conditions ($Y_{\rm e} \simeq 0.5$) being different from the case of domination by the Fe-peak nuclei with the largest binding energy per nucleon that have a proton to nucleon ratio close to the prescribed electron fraction. Particularly, the lower limit of the temperature in the NSE calculation is identified at about $2\times10^9 \rm K$. If the temperature is lower than this limit, the NSE solutions will not be reliable. Therefore, in our calculations we assume that all nuclear reactions would cease when the temperature is lower than this limit.

Thus NSE originates in the study of proton-rich state of matter, but it can be used in the description of all the state of matter. The limit of electron fraction, $Y_{\rm e} \lesssim 0.5$, has been canceled, and the more real state of matter can be described through these equations of NSE. The NSE code in proton-rich environments can be downloaded from \url{http://cococubed.asu.edu/code\_pages/nse.shtml}.

\subsection{Thermodynamics}\label{sec_thermodynamics}

The contributions to the pressure from degenerate electrons and from neutrinos should be included in the equation of state. It can be written as \beq\  p=p_{\rm gas}+p_{\rm rad}+p_{\rm e}+p_\nu.\eeq

The gas pressure from free nucleons $p_{\rm gas}$ is \beq\ p_{\rm gas}= \sum\limits_{j} n_j k_B T. \eeq

The disk is optical thick for the photons definitely, so photon radiation pressure $p_{\rm rad}$ is \beq\ p_{\rm rad}=aT^4/3. \eeq

The electron pressure $p_{\rm e}$ is from both electrons and positrons and should be calculated using the exact Fermi-Dirac distribution \citep[e.g.,][]{Chen2007,Liu2007}. No asymptotic expansions are valid because electrons are neither nondegenerate nor strongly degenerate, and they are not ultrarelativistic at all radii. It reads \beq\ p_{\rm e}=p_{\rm e^{-}}+ p_{\rm e^{+}}, \eeq with \beq\ p_{\rm e^{\mp}}= \frac{1}{3 {\pi}^2 {\hbar}^3 c^3} \int_0^{\infty} d p \frac{p^4}{\sqrt{p^2 c^2+{m_{\rm e}}^2 c^4}} \frac{1}{e^{({\sqrt{p^2 c^2+{m_{\rm e}}^2 c^4} \mp {\mu_{\rm e}})/k_{\rm B} T}}+1}. \eeq

\cite{Pan2012a,Pan2012b} calculated the one dimensional Boltzmann equation of the neutrino and anti-neutrino transport in accretion disks. From the solutions of NDAFs, the bridging formula valid in both the neutrino optically thin and thick regimes can also be used as a good approximation. We adopt the formula for the energy density of neutrinos ${u}_{\nu}$ \citep[e.g.,][]{Di Matteo2002,Kohri2005,Liu2007},\beq u_{\rm \nu}=\sum_{i} \frac{(7/8)a T^4 (\tau_{{\nu}_i}/2+1/ \sqrt{3})} {\tau_{{\nu}_i}/2+1/ \sqrt{3}+1/(3 \tau_{a,{\nu}_i})}. \eeq The neutrino pressure $p_{\rm \nu}$ is \beq\ p_{\rm \nu}=\frac{u_{\rm \nu}} {3}. \eeq

The cooling rate appearing in the equation (\ref{energy}) is composed of cooling rates of photodisintegration, neutrino emitting, and photon radiation \beq Q^{-}=Q_{\rm ph} +Q_{\nu}+Q_{\rm rad}. \eeq  However, the cooling due to the photon radiation is always much smaller than the other two so that $Q_{\rm rad}$ is ignored in our paper.

The cooling rate by photodisintegration, mainly related to $\alpha$-particles, can be written as $Q_{\rm ph}$ is \beq\ \ Q_{\rm ph} = 6.8 \times 10^{28} \rho_{10} V_r H \frac{d X_{\rm nuc}}{d r}~{\mathrm {cgs~units}}, \eeq where ${\rho}_{10} \equiv {\rho}/10^{10}{\rm g~cm^{-3}}$, and $X_{\rm nuc}$ is the mass fraction of free nucleons. \citep[e.g.,][]{Kohri2005,Liu2007}. Here we ignore cooling rate by disintegration of other heavy nuclei, because of the lower number density of these nuclei and the absolutely dominant advective cooling rate in the outer region.

The cooling rate due to neutrino loss $Q_\nu$ is expressed in accordance with the above equation of energy density of neutrinos \citep[e.g.,][]{Di Matteo2002,Kohri2005,Liu2007},\beq\ Q_{\nu}=\sum_{i} \frac{(7/8) {\sigma} T^4}{(3/4)[\tau_{{\nu}_i}/2+1/ \sqrt{3}+1/(3 \tau_{a,{\nu}_i})]}. \eeq

\subsection{Neutrino luminosity and neutrino annihilation luminosity}

The neutrino radiation luminosity can be calculated by integrating the neutrino cooling rate $Q_{\nu}$ along the disk, which is obtained as \beq L_{\rm \nu}=4 \pi \int_{{\rm max}(r_{\rm ms}, r_{\rm tr})}^{r_{\rm out}} Q_{\rm \nu} r d r, \eeq where $r_{\rm out}$ is the outer edge of the disk, which is fixed on $500$ Schwarzschild radius in our calculations; The lower bound is defined as the largest one of the neutrino trapping radius $r_{\rm tr}$ and the marginally stable radius of the black hole $r_{\rm ms}$, in order to take account the effect of neutrino trapping.

In the classic accretion theory, the radiation energy generated near the equatorial plane diffuses toward the disk surface at the speed of $\sim c/3 \tau$ \citep{Mihalas1984}, where $\tau$ is the total optical depth. Thus, the timescale of radiative diffusion is $t_{\rm diff}= H / (c/3 \tau)$ \citep[e.g.,][]{Ohsuga2002}. We get $V_n$ for neutrinos replace $c$ for photons, which is related to the energy of neutrino $\sim 3.7 k_B T$ \citep[e.g.,][]{Di Matteo2002,Liu2007,Liu2012a}. $V_n$ can be estimated by $\sim (3.7 k_B T c^2/0.07 \rm eV)^{1/2}$, where $0.07 \rm eV$ roughly equals to the low limit of neutrino rest-mass energy. Since the accretion timescale $t_{\rm acc}$ is given by $-r/{V_r}$, the condition in which the neutrino radiation energy is trapped in the flow and falls onto black hole is written as $t_{\rm diff}> t_{\rm acc}$. If only the electron neutrino optical depth has been considered, we can approximatively define the trapping radius \beq r_{\rm tr} \simeq -\frac{3 \tau_{\nu_{\rm e}}H V_r}{V_{\rm n}}. \eeq Obviously, the effect of neutrino trapping will greatly affect the annihilation luminosity.

For the calculation of the neutrino annihilation luminosity we follow the approach in \citet{Ruffert1997}, \citet{Popham1999}, \citet{Rosswog2003}, \citet{Liu2007} and \citet{Kawanaka2012b}. The disk is modeled as a grid of cells in the equatorial plane. A cell $k$ has its mean neutrino energy $\varepsilon_{\nu_i}^k$, neutrino radiation luminosity $l_{\nu_i}^k$, and distance to a space point above (or below) the disk $d_k$. $l_{\nu_i}^k$ can be calculated by the surface integral of cooling rate of each flavor of neutrino in the cell $k$ according the form of Equation (42) before summation. The angle at which neutrinos from cell $k$ encounter antineutrinos from another cell $k'$ at that point is denoted as $\theta_{kk'}$. Then the neutrino annihilation luminosity at that point is given by the summation over all pairs of cells, \beq l_{\nu \overline{\nu}}=\sum_{i} A_{1,i} \sum_k \frac{l_{\nu_i}^k}{d_k^2} \sum_{k'} \frac{l_{\overline{\nu}_i}^{k'}}{d_{k'}^2}(\varepsilon_{\nu_i}^k + \varepsilon_{\overline{\nu}_i}^{k'}) {(1-\cos {\theta_{kk'}})}^2 \nonumber\\+ \sum_{i} A_{2,i} \sum_k \frac{l_{\nu_i}^k}{d_k^2} \sum_{k'} \frac{l_{\overline{\nu}_i}^{k'}}{d_{k'}^2} \frac{\varepsilon_{\nu_i}^k +\varepsilon_{\overline{\nu}_i}^{k'}}{\varepsilon_{\nu_i}^k \varepsilon_{\overline{\nu}_i}^{k'}} {(1-\cos {\theta_{kk'}})},\eeq where $A_{1,i} = (1 / 12\pi^2)[\sigma_0/c{(m_{\rm e} c^2)}^2][{(C_{V,\nu_{i}}-C_{A,\nu_{i}})}^2 +{(C_{V,\nu_{i}}+C_{A,\nu_{i}})}^2]$ , $A_{2,i} = (1/6\pi^2) (\sigma_0/c)$ $ (2 C_{V,\nu_{i}}^2-C_{A,\nu_{i}}^2)$. The total neutrino annihilation luminosity is obtained by the integration over the whole space outside the black hole and the disk, \beq\ L_{\nu \overline{\nu}}=4 \pi \int_{{\rm max}(r_{\rm ms}, r_{\rm tr})}^\infty \int_H^\infty l_{\nu \overline{\nu}} r d r d z, \eeq where the inner edge is depended on the status of neutrino trapping \citep[e.g.,][]{Di Matteo2002,Liu2012a}.

\section{Numerical methods}\label{sec_num_meth}

To obtain the disk solutions, we have to solve the fundamental equations (\ref{continuity}), (\ref{energy}), (\ref{radial-momentum}), (\ref{angular-momentum}), (\ref{vertical-eq}), and (\ref{bridging-formula}) for the independent variables $\rho$, $T$, $V_r$, $\mathcal{L}$, $H$, and $\mu_{\rm e}$. In this paper, we follow \cite{Matsumoto1984} to use the shooting method for solving these equations, which is one of the popular methods for solving the boundary value problem of differential equations. However, there are two obstacles for solving these equations with shooting method. One is the instability of numerical integration and the other is the sonic point.

The first obstacle is due to the stiffness of these equations. It is numerically unstable to integrate the equations inwards with a certain explicit method (e.g. Runge-Kutta method). Fortunately, this integration instability can be overcome by the implicit integration. In this paper, we follow \cite{Matsumoto1984} to use the first order backward Euler method in shooting integration. In this method, for example, the differential equation $df/dr=g$ is approximated by the backward difference as $(f_{i+1}-f_i)/(r_{i+1}-r_i)=g_{i+1}$, where $g$ is an algebraic expression of $f$, and the value of $f$ on the grid point $r_i$ is known but unknown on $r_{i+1}$. With this kind of approximation, Equations (\ref{continuity}), (\ref{energy}), (\ref{radial-momentum}), (\ref{angular-momentum}), (\ref{vertical-eq}), and (\ref{bridging-formula}) are reduced to six nonlinear algebraic equations for the independent variables $\rho_{i+1}$, $T_{i+1}$, $V_{r,i+1}$, $\mathcal{L}_{i+1}$, $H_{i+1}$, and $\mu_{\rm{e}, i+1}$. In each integration step, we solve these equations with Newton-Raphson method. The initial guessed values of those independent variables on $r_{i+1}$ are set to be the determined values on $r_i$, which are either the results of the last step or the values on boundary in the first step. Even though the first order backward Euler scheme is less accurate than other alternative high-order difference scheme (e.g. central difference scheme), we found, in practice, it is a good method for overcoming the stiffness in our equations and has faster converging speed in the integration even than other high-order implicit difference scheme.

The other obstacle is due to the sonic point. As mentioned by many previous works \citep[e.g.][]{Sadowski2009,Matsumoto1984}, the derivative $d\ln V_r/d\ln r$ would tend to a $0/0$ form limit when the radial velocity tends to the local sound speed of the accreted gas. In fact, the numerical computations are performed in computer with a finite machine accuracy, so the sonic point cannot be really reached but only approached and spanned from the subsonic region to the supersonic region. In \cite{Matsumoto1984}, they presented a detailed and completed research on the mathematical and physical properties of the sonic points in viscous transonic flows. Benefiting from their research, we use their adaptive grid scheme to perform the shooting integration \citep[see the details in Appendix 2 of][]{Matsumoto1984}, which shows a robust ability on transonic integration in our practice.

After overcoming the above two obstacles, the shooting integration for solving our equations becomes possible. The angular momentum at the inner edge of the disk, $\mathcal{L}_{\rm{in}}$, becomes the eigenvalue of the shooting integration. At the beginning of the first shooting, we set the guessed value of $\mathcal{L}_{\rm{in}}$ to be the Keplerian angular momentum at the marginally stable orbit. If the guessed value is larger than the proper value $\mathcal{L}_{\rm{in,0}}$, the integration will be unable to converge near the sonic point. However, if the guessed value is less than $\mathcal{L}_{\rm{in,0}}$, a fully subsonic solution will be obtained. If the former case is met, the failed value will be used to update the upper limit of $\mathcal{L}_{\rm{in}}$, on the contrary, if the latter case occurs, the lower limit will be updated. After updating the upper and lower limits, a new shooting will begin with a different guessed $\mathcal{L}_{\rm{in}}$, whose value is set to be the midpoint of those limits. This kind of shooting integrations are repeated until a self-consistent transonic solution is obtained.

\section{Numerical results}\label{sec_num_res}

In our model for NDAF, there are four parameters, the viscous parameter $\alpha$, black hole mass $M$, dimensionless black hole spin $a_*$ ($\equiv a/M$), and dimensionless accretion rate $\dot m$ [$\equiv \dot M/(M_\odot~\rm s^{-1})$]. In order to concentrate on the more important effects of the different black hole spins and accretion rates, we fix the viscous parameter and black hole mass with the typical values of NDAF, $\alpha=0.1$ and $M = 3~M_\odot$, and investigate sixteen solutions with the different black hole spins ranged in $a_*$=0, 0.5, 0.9, 0.99 and accretion rates ranged in $\dot m$=0.03, 0.1, 1, 10. Where, the selected values of $a_*$ cover the cases with no, moderate, high and extreme spins of the black holes, which is an essential qualification for the emergence of jet breakout in collapsars, the plausible progenitor of long GRBs \citep[e.g.,][]{Nagakura2011,Nagakura2012}. Meanwhile, the selected values of $\dot m$ correspond to the cases with low, moderate, high and superhigh accretion rates. These solutions can be obtained by numerically solve the fundamental equations (\ref{continuity}), (\ref{energy}), (\ref{radial-momentum}), (\ref{angular-momentum}), (\ref{vertical-eq}), and (\ref{bridging-formula}) with the numerical method described in section \ref{sec_num_meth}.

\subsection{Structure}

\begin{figure}
\centering
\includegraphics[angle=0,scale=0.6]{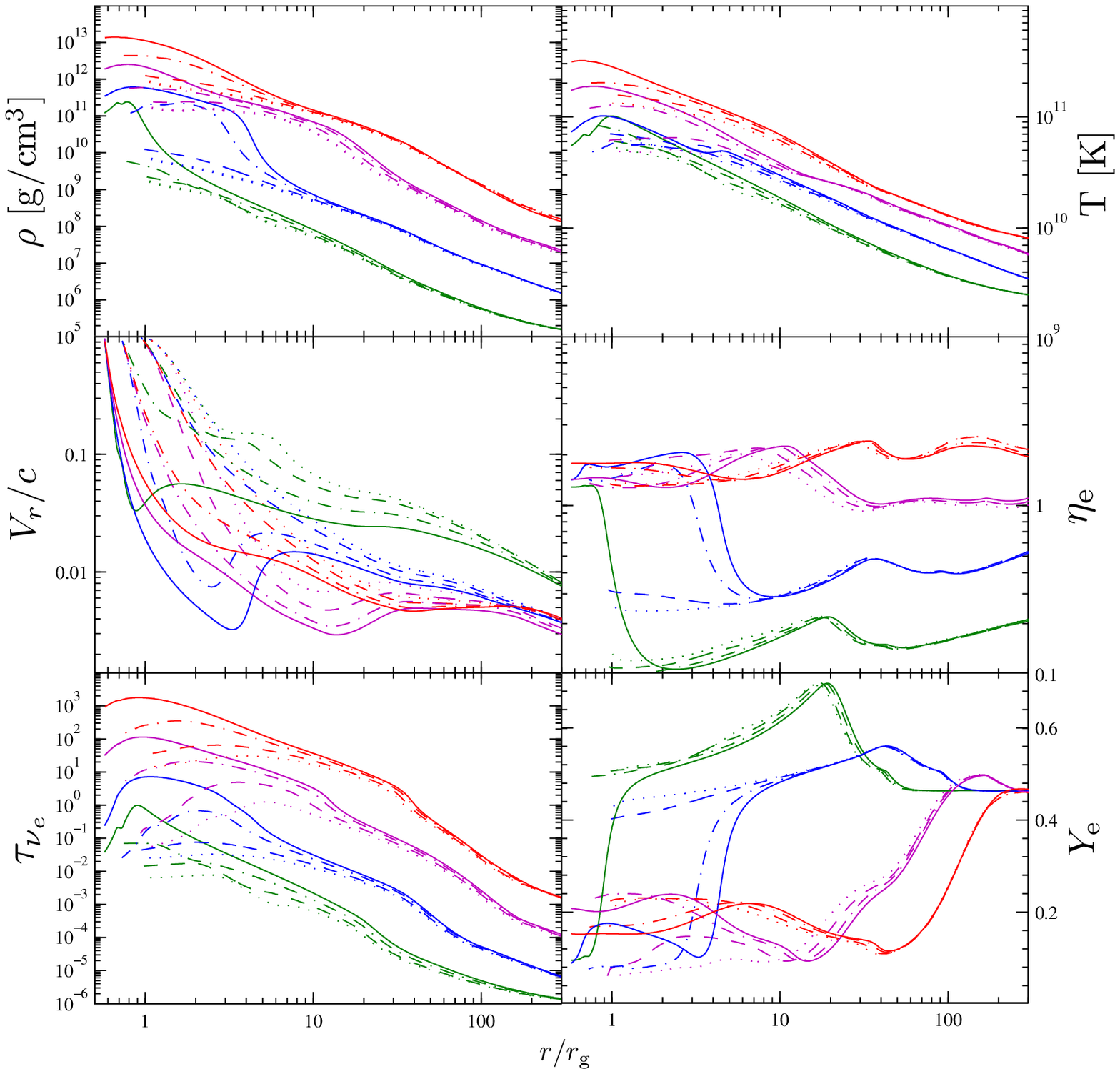}
\caption{The profiles for the sixteen solutions. The colors, green, blue, purple, and red denote the different accretion rates $\dot m$ = $0.03$, $0.1$, $1$, and $10$, respectively. The line styles, dotted, dashed, dot-dashed, and solid denote the different black hole spin $a_*$ = $0$, $0.5$, $0.9$, and $0.99$. The six panels show the profiles of density $\rho$, temperature $T$, radial velocity $V_r$, electron degeneracy $\eta_{\rm e}$, optical depth of electron neutrino $\tau_{\nu_{\rm e}}$, and electron fraction $Y_{\rm e}$ from left to right and upper to lower, respectively.}
\label{fig_struc}
\end{figure}

In Figure \ref{fig_struc}, we show the structures of the sixteen solutions for comparing them with each other. There are six panels and they correspond with the profiles of density $\rho$, temperature $T$, radial velocity $V_r$, electron degeneracy $\eta_{\rm e}$, optical depth of electron neutrino $\tau_{\nu_{\rm e}}$, and electron fraction $Y_{\rm e}$. They reveal different spans of variation that $\rho$ increases by about $6$ orders of magnitude, $T$ increases by about $1$ orders of magnitude, and $\tau_{\nu_{\rm e}}$ increases by about $5$ orders of magnitude from the outer to inner region, and, in the innermost region, $\rho$ reaches $\sim10^{13} \rm g~cm^{-3}$, $T$ reaches $\sim 10^{12}\rm K$, and $\tau_{\nu_{\rm e}}$ reaches $\sim 1000$, which are extremely dense, hot and neutrino-optically thick \citep[see, e.g.,][]{Li2013}. The difference between the effects of the accretion rate and the black hole spin is obvious. One can find that the profiles can be collected to four groups with the same color (the same accretion rate). It implies the effect of accretion rate is global. Meanwhile, the profiles with different line-styles (different black hole spins) in a colored group become more and more dispersive from outer to inner. It implies that the effect of black hole spin becomes remarkable only in the locations close to the black hole.

There is a remarkable feature in the profiles of the radial velocity. They all tend to the light speed $c$ in the inner region and the radii satisfied $V_r/c=1$ (the locations of the black hole horizon) only determined by the black hole spin. It proofs that our model and the calculations are all consistent with the general relativity.

The electron degeneracy is an important physical parameter that affects electron fraction, degeneracy pressure, and neutrino cooling \citep{Chen2007}. The profiles of $\eta_{\rm e}$, which represents the degree of electron degeneracy, in Figure \ref{fig_struc} is similar with the ones of \citet{Chen2007} and \citet{Kawanaka2007}. However, there is a little difference between the profiles of $Y_{\rm e}$ in Figure \ref{fig_struc} and the previous works \citep[e.g.,][]{Liu2007,Chen2007,Kawanaka2007}. Our $Y_{\rm e}$'s can become larger than $0.5$ but cannot in those previous works because of the different description of NSE we used. Our $Y_{\rm e}$'s all tend to $\sim 0.46$ at the outer boundary of the disk, which have a little different value $\sim 0.42$ in \citet{Kawanaka2007}.

\begin{figure*}
\includegraphics[angle=0,scale=0.6]{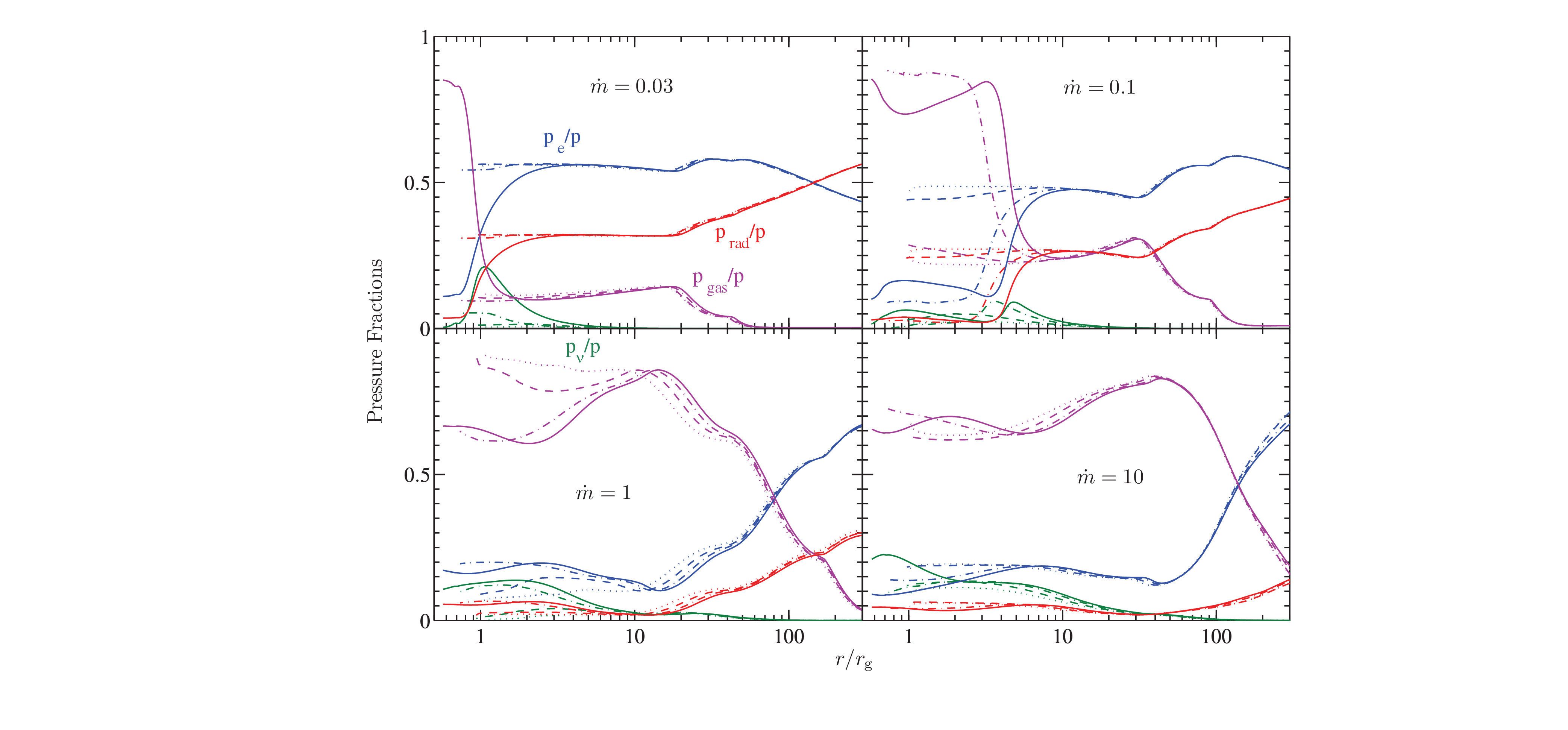}
\caption{Contributions to the total pressure $p$. Here, the colors, blue, red, purple, and green denote the different contributions from the degeneracy pressure of electrons $p_{\rm e}$, radiation pressure of photons $p_{\rm rad}$, gas pressure of nucleons $p_{\rm gas}$, and radiation pressure of neutrinos $p_{\nu}$ respectively. The meanings of the line-styles are the same with ones in Figure \ref{fig_struc}. The four panels correspond to the different accretion rates.}
\label{fig_press}
\end{figure*}

In Figure \ref{fig_press}, we show the contributions to the total pressure $p$ from the gas pressure of nucleons $p_{\rm gas}$, radiation pressure of photons $p_{\rm rad}$, degeneracy pressure of electrons $p_{\rm e}$ and radiation pressure of neutrinos $p_{\nu}$ respectively. Similar to the seeing in Figure \ref{fig_struc}, the effect of black hole spin is revealed by the dispersion of the profiles with different line-styles in a colored group. The more dispersion implies the more remarkable effects. One can see that the effects of black hole spins is still constrained in the inner region as seeing in Figure \ref{fig_struc}. However, a little detail revealed from Figure \ref{fig_press} should be mentioned here. The effects of black hole spins on the solutions with moderate ($\dot m=0.1$) and high ($\dot m=1$) accretion rates are more remarkable (more dispersive) than the solutions with low ($\dot m=0.03$) and superhigh ($\dot m=10$) accretion rates. The reasons are that the cooling of neutrino radiation in low accretion case is too weak to be sensitive to black hole spins only except for the solution with extremely spinning black hole, whereas the effect of black hole spin is damped by the superhigh accretion rate so that it is also insensitive to black hole spins in the case with superhigh accretion rate.

Focus on the profiles of $p_{\rm e}/p$ and $p_{\rm gas}/p$ in Figure \ref{fig_press}, one can see that the contributions of $p_{\rm gas}$ exceed $p_{\rm e}$ in some certain radii. The exceeding points only exist for the cases with high or extreme black hole spins in the solutions with low and moderate accretion rates, while they always exist in the solutions with high and superhigh accretion rates and their locations shift outwards for larger accretion rates. These are also due to the competitive relationship between the effects of the black hole spin and accretion rate on the neutrino cooling.

For the contributions of radiation pressure of neutrinos $p_{\nu}$, one can see that they are fully ignorable in outer region, which is always neutrino-optically thin, while they become notable and even exceed the photon radiation pressure in the inner region, which is neutrino-optically thick.

\begin{figure*}
\includegraphics[angle=0,scale=0.6]{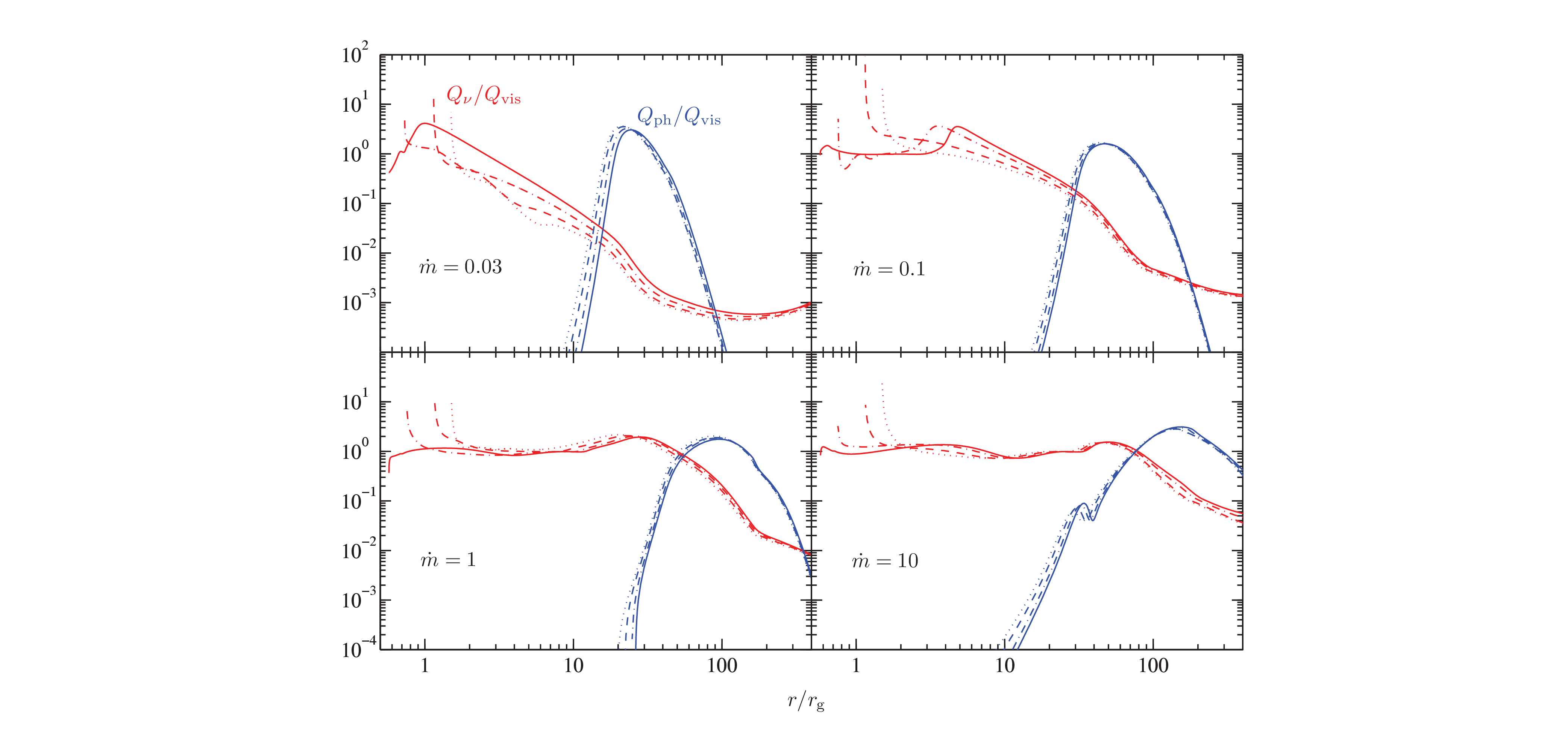}
\caption{Cooling rates normalized by the viscous heating rate $Q_{\rm vis}$. The upper four panels exhibit the normalized advection cooling rates $Q_{\rm adv}/Q_{\rm vis}$ and the lower four panels is for the normalized neutrino cooling rates $Q_{\nu}/Q_{\rm vis}$, and photodisintegration cooling rates $Q_{\rm ph}/Q_{\rm vis}$. The meanings of different line-styles are same with the ones in Fig. \ref{fig_struc} and Fig. \ref{fig_press}.}
\label{fig_coolings}
\end{figure*}

In Figure \ref{fig_coolings}, we show the cooling rates normalized by the viscous heating rate $Q_{\rm vis}$. The photon coolings in our solutions are always much lower than the other coolings, so we ignore it in this figure. The profiles of our $Q_{\rm adv}/Q_{\rm vis}$'s are very similar to the ones in \citet{Popham1999}. $Q_{\rm adv}/Q_{\rm vis}\sim 1$ in the outer region implies that the flow is advection dominated. The photodisintegration cooling causes the decrease of $Q_{\rm adv}/Q_{\rm vis}$'s even become negative in outer and middle regions. The neutrino cooling make $Q_{\rm adv}/Q_{\rm vis}$'s decrease to negative again in inner region. All of these behaviors are consistent with the relevant results of previous works \citep[e.g.,][]{Popham1999,Liu2007,Chen2007}. A special behavior of $Q_{\rm adv}/Q_{\rm vis}$'s and $Q_{\nu}/Q_{\rm vis}$'s that their values rapidly increase to much larger than unity near the inner edge of disks is due to the rapidly decrease of viscous heating $Q_{\rm vis}$ in the fast inflowing flows.

\begin{figure*}
\includegraphics[angle=0,scale=0.7]{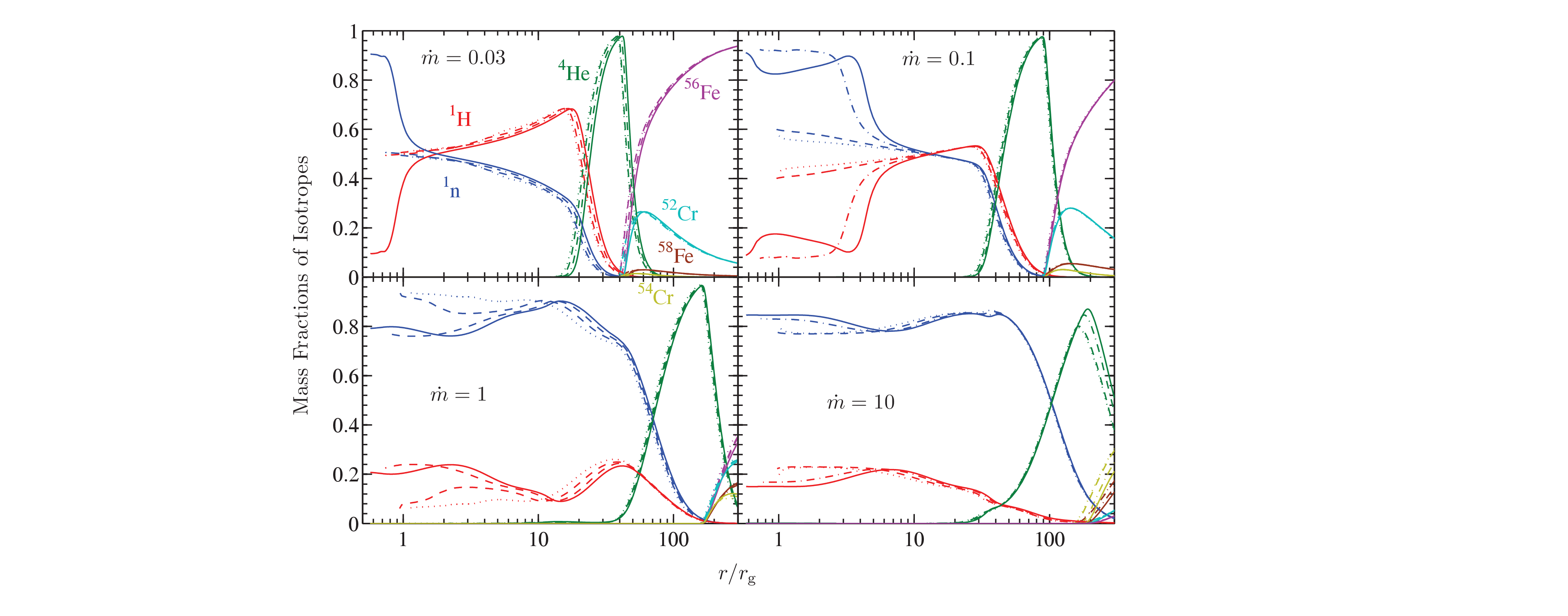}
\caption{The radial distributions of the mass fraction of seven major nucleons $\rm ^1n$, $\rm ^1H$, $\rm ^4He$, $\rm ^{52}Cr$, $\rm ^{54}Cr$, $\rm ^{56}Fe$. The meanings of different line-styles are same with the ones in Fig. \ref{fig_struc}, \ref{fig_press}, and \ref{fig_coolings}.}
\label{fig_nucleons}
\end{figure*}

Since we have considered the detailed nucleosynthesis in our model, we can obtain and trace the radial variation of more than 40 nucleons with our calculation.
Figure \ref{fig_nucleons} shows the radial distributions of the mass fractions of seven major nucleons $\rm ^1n$, $\rm ^1H$, $\rm ^4He$, $\rm ^{52}Cr$, $\rm ^{54}Cr$, $\rm ^{56}Fe$ and $\rm ^{58}Fe$, which cover almost $99\%$ mass of flow. The mass fraction of $\rm ^{56}Fe$ dominates in the outer region for all the accretion rate. In the middle region, $\rm ^4 He$ is dominant for all the accretion rate. Free neutrons and protons are dominant via photodisintegration in the inner region with the hot and dense state. The size of the region dominated by free nucleons is determined by the accretion rate. The spin of black hole is also affected on the proportion of free protons and neutrons in the inner region. Most of the free protons turn into the free neutrons due to the Urca process \citep[e.g.,][]{Liu2007,Kawanaka2007}, which causes the dominant free neutrons and the decrease of electron fraction. Comparing with \citet{Chen2007} and \citet{Liu2007}, the more heavy nuclei appeared in the outer region of the disk, the more possible structure and component distribution have been described. It is an implication for the origin of heavy nuclei in GRBs accounting for the detection of Fe K$\alpha$ X-ray lines and other emission lines \citep[e.g.,][]{Lazzati1999}, which can play an important role in understanding the nature of GRBs, especially its central engine. The neutron-rich NSE has been used in \citet{Kawanaka2007} (unfortunately, they did not show the distribution of heavy nuclei in the outer disk), in which electron fraction $Y_{\rm e}$ has been limited less than $0.5$. The reasonable NSE, we chose, is required that the range for $Y_{\rm e}$ is $[0,1]$, which can certify the solutions naturally and reasonably. Comparing with \citet{Liu2013}, kinds and distribution of elements are not different in the radial and vertical coordinates. $\rm ^{56}Ni$ dominates at the disk surface for lower accretion rate (e.g., $0.05~M_\odot$ $\rm s^{-1}$), and $\rm ^{56}Fe$ dominates for larger accretion rate (e.g., $1~M_\odot$ $\rm s^{-1}$), corresponding to $Y_{\rm e}$ around 0.49 and 0.47, respectively. In addition, according to Equation (29), the profiles of $Y_{\rm e}$ in Figure \ref{fig_struc} can be indicated by Figure \ref{fig_nucleons} if we considered that the mass fraction approximately equals the number density.

\subsection{Luminosity}

In this paper, we concern the energetic estimation for boosting GRB through the neutrino annihilation outside NDAF. Therefore, we only concern the total annihilation luminosity but not the exact distribution of the annihilation energy and we calculate the neutrino trapping radius and the annihilation luminosity without taking account of the general relativistic effects on the neutrino trajectory to avoid the complexity in this calculation. In Table \ref{tab1}, we list out the neutrino radiation luminosity $L_\nu$, neutrino annihilation luminosity $L_{\nu \bar{\nu}}$, efficiency of energy deposition $L_{\nu \bar{\nu}}/L_\nu$, neutrino trapping radius $r_{\rm tr}$, and the radius of marginally stable orbit $r_{\rm ms}$ of our sixteen solutions. The upper limit of neutrino luminosity reaches about $10^{55} \rm erg ~s^{-1}$, which is near the limit of the power of the Kerr black hole, for the solution with $\dot{m}=10$ and $a_*=0.99$. We notice that most of results about neutrino annihilation luminosity are higher than $10^{49}\rm{erg~s^{-1}}$ and thus are likely to be adequate for GRBs \citep{Zhang2011} even taking into account the effect of neutrino trapping, especially for the high accretion rate and rapidly spinning black hole.

Comparing with the work of \cite{Popham1999}, $L_{\nu \bar{\nu}}$ is much smaller than theirs for the solutions with superhigh accretion rate ($\dot m=10$) since we consider the effects of neutrino trapping. In particular, they had $L_{\nu \bar{\nu}}=2\times10^{53}\rm{erg/s}$ for $a_*=0$, $\dot m=10$, and $L_{\nu \bar{\nu}}=8.2\times10^{53}\rm{erg/s}$ for $a_*=0.5$, $\dot m=10$, while we obtain more reasonable values $L_{\nu \bar{\nu}}=2.94\times10^{52}\rm{erg/s}$, and $L_{\nu \bar{\nu}}=3.17\times10^{52}\rm{erg/s}$, respectively. This implies that the influence of neutrino trapping cannot be ignored especially for the superhigh accretion cases.

\cite{Zalamea2011} fully considered the general relativistic effects on the neutrino trajectory in their annihilation calculation. They compared their results with \cite{Popham1999}. They stated that $L_{\nu \bar{\nu}}$ would be overestimated 10 times by the calculation without taking account general relativistic effects. Thus, taking a conservative estimation, the 10 times lower $L_{\nu \bar{\nu}}$ listed in Table \ref{tab1} can be regarded as reasonable approximated values. Based on the results in Table \ref{tab1}, we approximate the neutrino radiation luminosity, annihilation luminosity and neutrino trapping radius with three analytic formulae as functions of black hole spin and accretion rate, \beq \log L_{\nu}(\rm{erg\ s^{-1}})&\approx& 52.5+1.17a_*+1.17\log\dot{m}, \label{fit_Lnu} \eeq \beq \log L_{\nu \bar{\nu}}(\rm{erg\ s^{-1}})&\approx& 49.5+2.45a_*+2.17\log\dot{m}, \label{fit_Lnunu} \eeq \beq r_{\rm tr}/r_g&\approx& -0.92+2.42a_*+5.95\log\dot{m}, \label{fit_rtr} \eeq where the negative value of $r_{\rm tr}$ predicted by equation (49) means there no trapping in the whole disk.

\section{Conclusions and discussion}\label{sec_con_dis}

In this paper, we calculated one-dimensional global solutions of NDAFs, taking into account strict Kerr metric, particular neutrino physics and precise nucleosynthesis processes, and discussed the structure and luminosity of NDAFs. \iffalse We find that the gas pressure of nucleons and neutrino cooling dominate in the inner region of the disk for large accretion rate, $\gtrsim 0.1~M_\odot$ $\rm s^{-1}$. \fi The electron degeneracy has significant effects in NDAFs and the electron fraction is about $0.46$ in the outer region. From the perspective of the mass fraction, free nucleons, $\rm ^4He$, and $\rm ^{56}Fe$ dominate in the inner, middle, and outer region, respectively. The influence of neutrino trapping on the annihilation is of importance for the superhigh accretion ($\dot M=10M_{\odot}~\rm s^{-1}$) and most of the sixteen solutions have an adequate annihilation luminosity for GRBs.

The inner region of NDAFs may be dynamically unstable \citep[e.g.,][]{Janiuk2007,Kawanaka2012a}. Time-dependent NDAFs should be calculated replacing the steady solutions to verify the instability of the disk. Time-dependent accretion disks around Kerr black holes has been investigated in \citet{Xue2011}, which can be as the basis for the study of time-dependent NDAF model.

Jet emission is an essential qualification of GRB events. Some jet emission mechanisms have been discussed in literatures. Basing the magnetic extraction of the rotational energy of a spinning black hole \citep{Blandford1977}, \citet{Di Matteo2002} and \citet{Kawanaka2012b} estimated the BZ-luminosity from NDAFs to budget the energy for jets and relevant GRBs. Recently, \citet{Yuan2012} presented an alternative magnetohydrodynamic mechanism for the emission of episodic jets, which also can be used to power GRBs. Without magnetic field, pair creation by neutrino annihilation outside  NDAFs \citep{Eichler1989}, in fact, also has ability for the jet emission and it may possess an additional virtue on low baryonic contamination at the jet ejection point. Our work in this paper belongs to the type of pair creation but we do not exploit this problem deeply due to avoiding the unnecessary complexity in the calculation of neutrino annihilation. Thus, a further work of us is to fully relativistically calculate the neutrino annihilation and the neutrino trapping radius, and obtain the the spacial distribution of energy deposition for our disk model like the works of \cite{Zalamea2011} \citep[also see, e.g.,][]{Birkl2007,Kovacs2011a,Kovacs2011b}. Even, it may be also attractive if we can combine the pair creation with the other magnetic mechanisms \citep[e.g.][]{Blandford1977,Yuan2012}, which seem not conflict with each other.

\acknowledgments
We thank the anonymous referee for instructive comments and helpful suggestions. This work was supported by the National Basic Research Program (973 Program) of China under Grant 2009CB824800, the National Natural Science Foundation of China under grants 11003016, 11073015, 11103015, 11222328, and 11233006.

\begin{figure*}
\includegraphics[angle=0,scale=0.65]{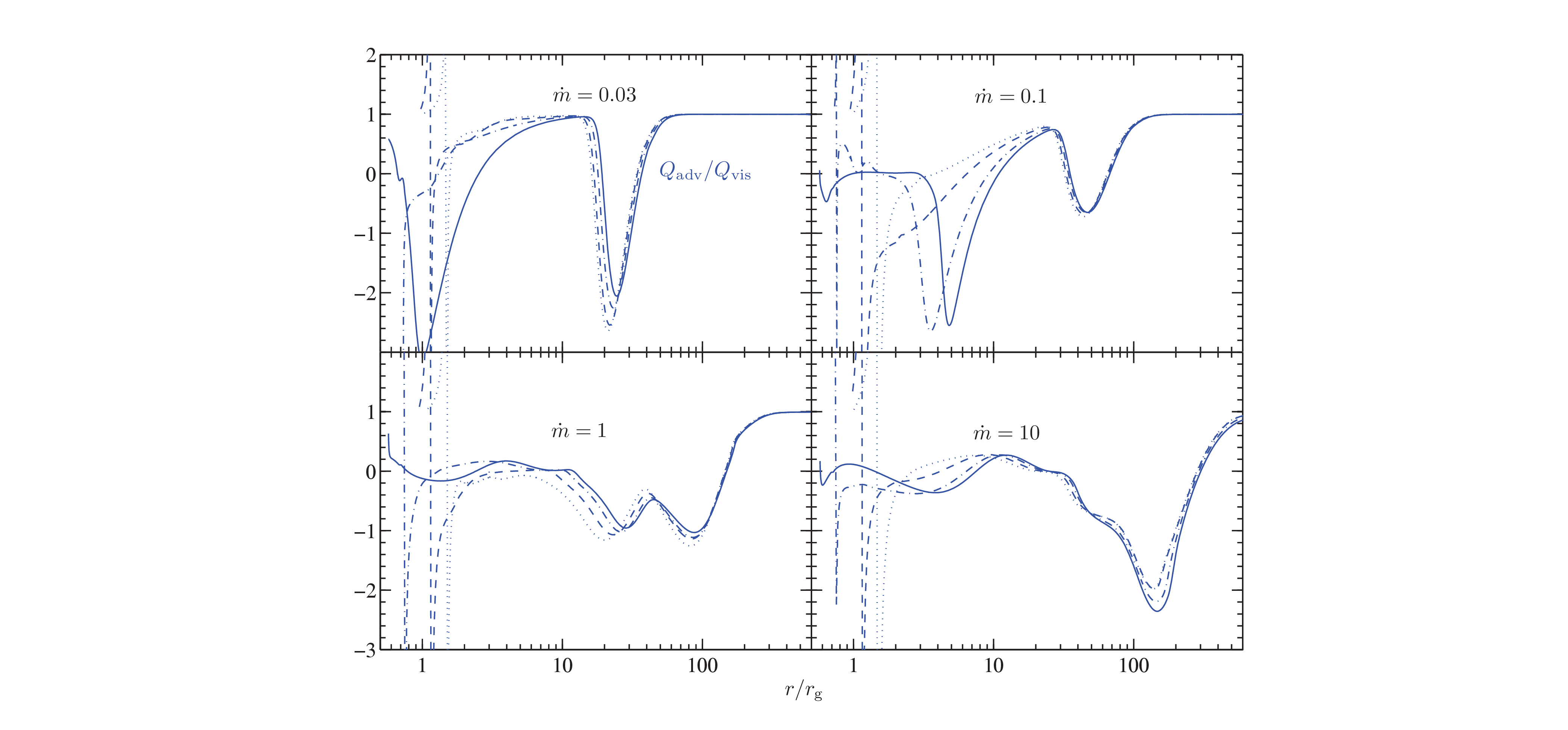}
\end{figure*}

\clearpage

\begin{deluxetable}{ccccccc}
\tabletypesize{\scriptsize}
%\rotate
\tablecaption{Power of NDAFs\label{tab1}}
\tablewidth{0pt}
\tablehead{
\colhead{$a_*$} & \colhead{$\dot{M}\ [M_{\odot}\ \rm{s^{-1}}]$} & \colhead{$L_\nu$ [$\rm{erg~ s^{-1}}$]
} & \colhead{$L_{\nu \bar{\nu}}$ [$\rm{erg~ s^{-1}}$]
} & \colhead{$L_{\nu \bar{\nu}}/L_\nu$} & \colhead{$r_{\rm{tr}}\ [r_{\rm{g}}]$}\tablenotemark{1} & \colhead{$r_{\rm{ms}}\ [r_{\rm{g}}]$\tablenotemark{2}}
}
\startdata
      0.00   &     0.03    & $2.04\times10^{50}$ & $1.72\times10^{44}$ & $8.43\times10^{-7}$ &    $<r_{\rm ms}$   &    3.000  \\
      0.00   &     0.10    & $4.58\times10^{51}$ & $9.11\times10^{47}$ & $1.99\times10^{-4}$ &    $<r_{\rm ms}$   &    3.000  \\
      0.00   &     1.00    & $1.00\times10^{53}$ & $8.82\times10^{50}$ & $8.82\times10^{-3}$ &    $<r_{\rm ms}$   &    3.000  \\
      0.00   &    10.00    & $7.60\times10^{53}$ & $2.94\times10^{52}$ & $3.87\times10^{-2}$ &    5.102   &    3.000  \\
      0.50   &     0.03    & $6.09\times10^{50}$ & $4.91\times10^{45}$ & $8.06\times10^{-6}$ &    $<r_{\rm ms}$   &    2.117  \\
      0.50   &     0.10    & $1.15\times10^{52}$ & $9.47\times10^{48}$ & $8.23\times10^{-4}$ &    $<r_{\rm ms}$   &    2.117  \\
      0.50   &     1.00    & $1.52\times10^{53}$ & $2.61\times10^{51}$ & $1.72\times10^{-2}$ &    $<r_{\rm ms}$   &    2.117  \\
      0.50   &    10.00    & $1.20\times10^{54}$ & $3.17\times10^{52}$ & $2.64\times10^{-2}$ &    6.095   &    2.117  \\
      0.90   &     0.03    & $3.29\times10^{51}$ & $3.08\times10^{47}$ & $9.36\times10^{-5}$ &    $<r_{\rm ms}$   &    1.160  \\
      0.90   &     0.10    & $3.45\times10^{52}$ & $1.79\times10^{50}$ & $5.19\times10^{-3}$ &    $<r_{\rm ms}$   &    1.160  \\
      0.90   &     1.00    & $3.04\times10^{53}$ & $1.76\times10^{52}$ & $5.79\times10^{-2}$ &    $<r_{\rm ms}$   &    1.160  \\
      0.90   &    10.00    & $3.10\times10^{54}$ & $3.73\times10^{52}$ & $1.20\times10^{-2}$ &    7.078   &    1.160  \\
      0.99   &     0.03    & $3.11\times10^{52}$ & $8.74\times10^{49}$ & $2.81\times10^{-3}$ &    $<r_{\rm ms}$   &    0.727  \\
      0.99   &     0.10    & $7.18\times10^{52}$ & $9.30\times10^{50}$ & $1.30\times10^{-2}$ &    $<r_{\rm ms}$   &    0.727  \\
      0.99   &     1.00    & $6.92\times10^{53}$ & $5.08\times10^{52}$ & $7.34\times10^{-2}$ &    1.473   &    0.727  \\
      0.99   &    10.00    & $6.38\times10^{54}$ & $4.19\times10^{52}$ & $6.57\times10^{-3}$ &    7.600   &    0.727  \\
\enddata
\tablenotetext{1}{$r_{\rm{tr}}$ is the neutrino trapping radius. $r_{\rm{g}}=2M$ is the Schwarzschild radius.}
\tablenotetext{2}{$r_{\rm{ms}}$ is the radius of marginally stable orbit.}
\end{deluxetable}

\clearpage

\clearpage

\begin{thebibliography}{99}
 \bibitem[Abramowicz et al.(1996)] {Abramowicz1996} Abramowicz, M. A., Chen, X.-M., Granath, M. \& Lasota J.-P. 1996, \apj, 471, 762
 \bibitem[Abramowicz et al.(1997)] {Abramowicz1997} Abramowicz, M. A., Lanza, A., \& Percival, M. J. 1997, \apj, 479, 179
 \bibitem[Birkl et al.(2007)]{Birkl2007} Birkl, R., Aloy, M.~A., Janka, H.-T., \& M\"{u}ller, E.\ 2007, \aap, 463, 51
 \bibitem[Blandford \& Znajek(1977)]{Blandford1977} Blandford, R.~D., \& Znajek, R.~L.\ 1977, \mnras, 179, 433
 \bibitem[Burrows \& Thompson(2004)]{Burrow2004} Burrows, A., \& Thompson, T. A. 2004, in Stellar Collapse, ed. C. L. Fryer (Dordrecht: Kluwer), 133
 \bibitem[Chen \& Beloborodov(2007)]{Chen2007} Chen, W.-X., \& Beloborodov, A.~M.\ 2007, \apj, 657, 383
 \bibitem[Di Matteo et al.(2002)]{Di Matteo2002} Di Matteo, T., Perna, R., \& Narayan, R.\ 2002, \apj, 579, 706
 \bibitem[Eichler et al.(1989)] {Eichler1989} Eichler, D., Livio, M., Piran, T., \& Schramm, D. N. 1989, Nature, 340, 126
 \bibitem[Gu et al.(2006)]{Gu2006} Gu, W.-M., Liu, T., \& Lu, J.-F.\ 2006, \apjl, 643, L87
 \bibitem[Itoh et al.(1989)]{Itoh1989} Itoh, N., Adachi, T., Nakagawa, M., Kohyama, Y., \& Munakata, H. 1989, \apj, 339, 354 (erratum 360, 741 [1990])
 \bibitem[Itoh et al.(1996)]{Itoh1996} Itoh, N., Hayashi, H., Nishikawa, A., \& Kohyama, Y.\ 1996, \apjs, 102, 411
 \bibitem[Janiuk et al.(2007)]{Janiuk2007} Janiuk, A., Yuan, Y., Perna, R., \& Di Matteo, T.\ 2007, \apj, 664, 1011
 \bibitem[Kato et al.(2008)]{Kato2008} Kato, S., Fukue, J., \& Mineshige, S. 2008, Black-Hole Accretion Disks: Towards a New Paradigm (Kyoto: Kyoto Univ. Press)
 \bibitem[Kawanaka \& Kohri(2012)]{Kawanaka2012a} Kawanaka, N., \& Kohri, K.\ 2012, \mnras, 419, 713
 \bibitem[Kawanaka \& Mineshige(2007)]{Kawanaka2007} Kawanaka, N., \& Mineshige, S.\ 2007, \apj, 662, 1156
 \bibitem[Kawanaka et al.(2012)]{Kawanaka2012b} Kawanaka, N., Piran, T., \& Krolik, J.~H.\ 2013, \apj, 766, 31
 \bibitem[Kohri \& Mineshige(2002)]{Kohri2002} Kohri, K., \& Mineshige, S.\ 2002, \apj, 577, 311
 \bibitem[Kohri et al.(2005)]{Kohri2005} Kohri, K., Narayan, R., \& Piran, T.\ 2005, \apj, 629, 341
 \bibitem[Kov{\'a}cs et al.(2011a)]{Kovacs2011a} Kov{\'a}cs, Z., Cheng, K.~S., \& Harko, T.\ 2011a, \mnras, 411, 1503
 \bibitem[Kov{\'a}cs \& Harko(2011b)]{Kovacs2011b} Kov{\'a}cs, Z., \& Harko, T.\ 2011b, \mnras, 417, 2330
 \bibitem[Lattimer \& Swesty(1991)]{Lattimer1991} Lattimer, J. M., \& Swesty, D. F.\ 1991, Nucl. Phys. A, 535, 331
 \bibitem[Lazzati et al.(1999)]{Lazzati1999} Lazzati, D., Campana, S., \& Ghisellini, G. 1999, \mnras, 304, L31
 \bibitem[Lee et al.(2005)]{Lee2005} Lee, W.~H., Ramirez-Ruiz, E., \& Page, D.\ 2005, \apj, 632, 421
 \bibitem[Lei et al.(2009)]{Lei2009} Lei, W.~H., Wang, D.~X., Zhang, L., et al.\ 2009, \apj, 700, 1970
 \bibitem[Li \& Liu(2013)]{Li2013} Li, A., \& Liu, T.\ 2013, \aap, accepted (arXiv:1305.2530)
 \bibitem[Liu et al.(2010a)]{Liu2010a} Liu, T., Gu, W.-M., Dai, Z.-G., \& Lu, J.-F.\ 2010a, \apj, 709, 851
 \bibitem[Liu et al.(2007)]{Liu2007} Liu, T., Gu, W.-M., Xue, L., \& Lu, J.-F.\ 2007, \apj, 661, 1025
 \bibitem[Liu et al.(2012a)]{Liu2012a} Liu, T., Gu, W.-M., Xue, L., \& Lu, J.-F.\ 2012a, \apss, 337, 711
 \bibitem[Liu et al.(2008)]{Liu2008} Liu, T., Gu, W.-M., Xue, L., Weng, S.-S., \& Lu, J.-F.\ 2008, \apj, 676, 545
 \bibitem[Liu et al.(2010b)]{Liu2010b} Liu, T., Liang, E.-W., Gu, W.-M., et al.\ 2010b, \aap, 516, A16
 \bibitem[Liu et al.(2012b)]{Liu2012b} Liu, T., Liang, E.-W., Gu, W.-M., et al.\ 2012b, \apj, 760, 63
 \bibitem[Liu et al.(2013)]{Liu2013} Liu, T., Xue, L., Gu, W.-M., \& Lu, J.-F.\ 2013, \apj, 762, 102
 \bibitem[Matsumoto et al.(1984)] {Matsumoto1984} Matsumoto, R., Kato, S., Fukue, J., \& Okazaki, A. T.,\ 1984, \pasj, 36, 71
 \bibitem[Mihalas \& Weibel Mihalas(1984)]{Mihalas1984} Mihalas, D., \& Weibel Mihalas, B.\ 1984, New York: Oxford University Press
 \bibitem[Nagakura(2012)]{Nagakura2012} Nagakura, H.\ 2013, \apj, 764, 139
 \bibitem[Nagakura et al.(2011)]{Nagakura2011} Nagakura, H., Ito, H., Kiuchi, K., \& Yamada, S.\ 2011, \apj, 731, 80
 \bibitem[Narayan et al.(1992)]{Narayan1992} Narayan, R., Paczynski, B., \& Piran, T.\ 1992, \apj, 395, L83
 \bibitem[Narayan et al.(2001)]{Narayan2001} Narayan, R., Piran, T., \& Kumar, P.\ 2001, \apj, 557, 949
 \bibitem[Narayan \& Yi(1995)]{Narayan1995} Narayan, R., \& Yi, I.\ 1995, \apj, 444, 231
 \bibitem[Ohsuga et al.(2002)]{Ohsuga2002} Ohsuga, K., Mineshige, S., Mori, M., \& Umemura, M.\ 2002, \apj, 574, 315
 \bibitem[Paczy\'{n}ski \& Wiita(1980)]{Paczynski1980} Paczy\'{n}ski, B., \& Wiita, P. J. 1980, \aap, 88, 23
 \bibitem[Pan \& Yuan(2012a)]{Pan2012a} Pan, Z., \& Yuan, Y.-F.\ 2012a, \apj, 759, 82
 \bibitem[Pan \& Yuan(2012b)]{Pan2012b} Pan, Z., \& Yuan, Y.-F.\ 2012b, \prd, 85, 064004
 \bibitem[Popham et al.(1999)]{Popham1999} Popham, R., Woosley, S.~E., \& Fryer, C.\ 1999, \apj, 518, 356
 \bibitem[Romero et al.(2010)]{Romero2010} Romero, G. E., Reynoso, M. M., \& Christiansen, H. R.\ 2010, \aap, 524, A4
 \bibitem[Rosswog et al.(2003)]{Rosswog2003} Rosswog, S., Ramirez-Ruiz, E., \& Davies, M. B. 2003, \mnras, 345, 1077
 \bibitem[Ruffert et al.(1999)]{Ruffert1997} Ruffert, M., Janka, H.-T., Takahashi, K., \& Sch\"{a}fer, G. 1997, \aap, 319, 122
 \bibitem[S\k{a}dowski(2009)] {Sadowski2009} S{\k a}dowski, A. 2009, \apjs, 183, 171
 \bibitem[Seitenzahl et al.(2008)]{Seitenzahl2008} Seitenzahl, I.~R., Timmes, F.~X., Marin-Lafl{\`e}che, A., et al.\ 2008, \apjl, 685, L129
 \bibitem[Sun et al.(2012)]{Sun2012} Sun, M.-Y., Liu, T., Gu, W.-M., \& Lu, J.-F.\ 2012, \apj, 752, 31
 \bibitem[Xue et al.(2011)]{Xue2011} Xue, L., S{\c a}dowski, A., Abramowicz, M.~A., \& Lu, J.-F.\ 2011, \apjs, 195, 7
 \bibitem[Yakovlev et al.(2001)]{Yakovlev2001}Yakovlev, D. G., Kaminker, A. D., Gnedin, O. Y., \& Haensel, P. 2001, Phys. Rep., 354, 1
 \bibitem[Yuan \& Zhang(2012)]{Yuan2012} Yuan, F., \& Zhang, B.\ 2012, \apj, 757, 56
 \bibitem[Yuan(2005)]{Yuan2005} Yuan, Y.-F. 2005, \prd, 72, 013007
 \bibitem[Zalamea \& Beloborodov(2011)]{Zalamea2011} Zalamea, I., \& Beloborodov, A.~M.\ 2011, \mnras, 410, 2302
 \bibitem[Zhang(2011)] {Zhang2011} Zhang, B. 2011, C. R. Phys., 12, 206
\end{thebibliography}
\end{document}